\documentclass[12pt]{aastex}
\def\etal{{et~al.\,}}

\def\teff{T$_{\rm eff}\,$}
\def\teffs{T$_{\rm eff}$s$\ $}
\def\Dwa{$\,$\uppercase\expandafter{\romannumeral5}$\,$}
\def\mic{$\mu$m$\,$}

\def\sles{\lower2pt\hbox{$\buildrel {\scriptstyle <}
   \over {\scriptstyle\sim}$}}

\def\sgreat{\lower2pt\hbox{$\buildrel {\scriptstyle >}
   \over {\scriptstyle\sim}$}}
\def\sharpnull#1{}

\begin{document}

\slugcomment{\bf}
\slugcomment{Accepted to Ap.J.}

\title{New CrH Opacities for the Study of L and Brown Dwarf Atmospheres}

\author{Adam Burrows\altaffilmark{1}, R.S. Ram\altaffilmark{2}, Peter
Bernath\altaffilmark{1,3},
C.M. Sharp\altaffilmark{1},
and J. A. Milsom\altaffilmark{4}}

\altaffiltext{1}{Department of Astronomy and Steward Observatory,
                 The University of Arizona, Tucson, AZ \ 85721;
                 burrows@jupiter.as.arizona.edu, csharp@as.arizona.edu}
\altaffiltext{2}{Department of Chemistry,
                 The University of Arizona, Tucson, AZ \ 85721;
                 rram@u.arizona.edu}
\altaffiltext{3}{Department of Chemistry, University of Waterloo,
Waterloo, Ontario, Canada  N2L 3G1; bernath@uwaterloo.ca}
\altaffiltext{4}{Department of Physics, The University of Arizona, Tucson,
AZ 85721;
                 milsom@physics.arizona.edu}

\begin{abstract}

In this paper, we calculate new line lists and opacities for the 12 bands of the A$^6\Sigma^{+}$ -- X$^6\Sigma^{+}$
transitions of the CrH molecule.  Identified in objects of the new L dwarf spectroscopic
class (many of which are brown dwarfs), as well as in sunspots, the CrH
molecule plays an important role in the diagnosis of low-temperature atmospheres.
As a tentative first application of these opacities, we employ our 
new theoretical CrH data in an atmospheres code to obtain a CrH/H$_2$ number ratio
for the skin of the L5 dwarf 2MASSI J1507038-151648 of $\sim 2-4\times 10^{-9}$,
in rough agreement with chemical equilibrium expectations.  Since in previous 
compilations the oscillator strength was off by more than an order of magnitude,
this agreement represents a modest advance.  However, in order to
fit the CrH abundance in the L dwarf spectral class, silicate clouds
need to be incorporated into the model.  Given that this subject is
still in a primitive stage of development, one should view any spectral model
in the L dwarf range as merely tentative.  Nevertheless, a necessary first
step in L dwarf modeling is a reliable CrH opacity algorithm and this is 
what we have here attempted to provide.

\end{abstract}


\section{Introduction}
\label{intro}

The CrH molecule had previously been identified 
in the spectra of sunspots (Engvold, W\"ohl, and Brault 1980) and 
S-type stars (Lindgren and Olofsson 1980), but recently CrH bands have been 
identified in the spectra of very-low-mass stars and brown dwarfs.  In fact, 
the 1-0 and 0-0 bands of the A$^6\Sigma^{+}$ -- X$^6\Sigma^{+}$
transition of CrH are now used as primary markers for the 
new L dwarf spectral class (Kirkpatrick et al. 1999a,b).
Hence, accurate CrH line lists and oscillator strengths 
are needed to calculate the CrH opacities now used to model the spectral energy
distributions of these transitional and substellar objects. The ground
X$^6\Sigma^{+}$ state of CrH is well
characterized from far-infrared (Corkery et al. 1991) and mid-infrared
(Lipus, Bachem, and Urban 1991) measurements.  However, until 
recently the same could not be said for the A$^6\Sigma^{+}$ state.

In this paper, we describe the recent spectroscopy of CrH and
constraining laboratory measurements that now allow us to calculate new 
opacities due to the A$^6\Sigma^{+}$ -- X$^6\Sigma^{+}$ transitions of the CrH molecule.
In \S\ref{spectrocrh}, we discuss the spectroscopy of the CrH molecule.
In \S\ref{generation}, we present the procedures and methods by which the new line lists were generated.
In \S\ref{opac}, we show how we calculate CrH opacities at a given temperature/pressure ($T/P$) point and 
in \S\ref{opacplots} we present representative opacity plots for CrH at temperatures and pressures
characteristic of L and brown dwarf atmospheres.  
In \S\ref{bdspec}, we provide and describe a few theoretical spectra generated with the new CrH opacity data
and derive an approximate CrH abundance in a representative L5 dwarf atmosphere.  
Though the focus of this paper is on the derivation of the new CrH line lists and opacities, we 
use these exploratory synthetic spectra to speculate on various features
of observed L dwarf spectra.  We plan to develop our findings in a subsequent paper.

\section{Spectroscopy of CrH}   
\label{spectrocrh}

Although the CrH molecule has been known since 1937 (Gaydon and Pearse 1937),
its electronic spectra were poorly characterized until 1993, 
when Ram, Jarman, and Bernath (1993) performed a rotational
analysis of the 0-0 band of the A$^6\Sigma^{+}$ -- X$^6\Sigma^{+}$
transition and obtained improved rotational constants for the $v^\prime$=0 vibrational
level of the A$^6\Sigma^{+}$ state. From the observed perturbations in the
A$^6\Sigma^{+}$ -- X$^6\Sigma^{+}$ 0-0 band, these authors also provided evidence of a low-lying
a$^4\Sigma^{+}$ state near 11186 cm$^{-1}$.  Recently, the 1-0 and 1-1 bands of the
A$^6\Sigma^{+}$ -- X$^6\Sigma^{+}$ transitions were remeasured and an improved set of constants were 
obtained for the $v^\prime$=1 vibrational level of the A$^6\Sigma^{+}$ state
(Bauschlicher et al. 2001). In that work
the band origin of the 2-0 band was determined by fitting the band head
position using extrapolated rotational constants for the $v^\prime$=2 level.
In the ground state all of the available data for $v^{\prime\prime}$=0, 1, and 2,
including the hyperfine-free pure rotational transitions
and the hyperfine-free
vibration-rotation line positions calculated using the constants of Lipus
et al. (1991), were used to obtain an improved set of spectroscopic
constants for CrH.  This work provided a set of much improved equilibrium vibrational
and rotational constants for the A$^6\Sigma^{+}$ state. The new experimental
values are in excellent agreement with the results of an extensive set of ab
initio calculations (Bauschlicher et al. 2001).  However, no experimental
measurements are available of the radiative lifetime of the excited A$^6\Sigma^{+}$
state.  These data are needed to compute oscillator strengths used to derive molecular
opacities.  Bauschlicher et al. (2001), therefore, computed the ab initio
transition dipole moment function for the A$^6\Sigma^{+}$ -- X$^6\Sigma^{+}$
transition.  They also computed the Einstein A constants for many
vibrational bands.  The estimated accuracy for these computed transition probabilities
is 10-20\%.   In \S\ref{opac}, we use these Einstein A coefficients to derive
opacities for the CrH molecule as a function of temperature and pressure.

\section{The Generation of the Line Lists for the A$^6\Sigma^{+}$ --
X$^6\Sigma^{+}$ transition of CrH}   
\label{generation}

     In order to calculate the molecular opacities of CrH, 
the ground state term values for the $v$ =0, 1, and 2
vibrational levels were calculated up to $J$ = 39.5 using the spectroscopic
constants reported in the previous paper (Bauschlicher et al. 2001). The
term values for the $v$=3 vibrational level were also calculated using the extrapolated spectroscopic
constants of the ground state. The transition wavenumbers for the main
branches R$_{11}$, R$_{22}$, R$_{33}$, R$_{44}$, R$_{55}$, R$_{66}$,
P$_{11}$, P$_{22}$, P$_{33}$, P$_{44}$, P$_{55}$ and P$_{66}$ were
taken from the available observations over the observed range of $J$ values.
For high $J$ lines beyond the range of our observations, the associated
wavenumbers were taken from the predictions based on the constants for the ground and excited
states.  The excited state term values for the observed bands were
calculated by combining the lower state term values with the transition
wavenumbers of the observed main branches. The high $J$ excited state term
values, for which main branches were not observed experimentally, were taken
from the predictions based on the spectroscopic constants for the excited
states.  Accurate excited state term values are necessary to predict the
line positions in the satellite branches, which were not observed in the
spectra because of their very weak intensity. The transition wavenumbers for
the satellite branches were calculated simply by taking the difference of
the excited and the ground state term values based on the selection rules.
In total, line positions of the 18 allowed satellite branches, namely,
P$_{31}$, P$_{42}$, P$_{53}$, P$_{64}$, Q$_{12}$, Q$_{21}$, Q$_{23}$, Q$_{32}$,
Q$_{34}$, Q$_{43}$, Q$_{45}$, Q$_{54}$, Q$_{56}$, Q$_{65}$, R$_{13}$, R$_{24}$,
R$_{35}$ and R$_{46}$ were calculated for the molecular opacity calculation.

The line strengths of various transitions were evaluated by computing
($A_{v^{\prime}-v^{\prime\prime}}$
$\times$ HLF)/(2$J^\prime$+1)) values in the branches of the different
bands.  Here the $A_{v^{\prime}-v^{\prime\prime}}$  are
the Einstein A values for the different bands considered (Bauschlicher et
al. 2001) and HLF are the H\"onl-London factors of the different branches.  
The H\"onl-London factors were derived by M. Dulick of the National Solar
Observatory (M. Dulick, private communication) and, except for a typographical error,
agree with the expressions in Kovacs (1969). The line positions and line
strengths were calculated for the 0-0, 0-1, 0-2, 0-3, 1-0, 1-1, 1-2, 1-3, 2-0, 
2-1, 2-2 and 2-3 bands of the  A$^6\Sigma^{+}$ -- X$^6\Sigma^{+}$ system of
CrH, where by convention $v^{\prime}$-$v^{\prime\prime}$ ($=\Delta{v}$) refers to
the upper and lower vibrational quantum numbers.  For bands
involving $v^\prime$ = 2, the excited state constants were obtained from an
extrapolation based on the experimental constants for the $v$ = 0 and 1
vibrational levels.

We arrange the computed sets of line strengths and line positions for
12 bands of CrH belonging to the A$^6\Sigma^{+}$ -- X$^6\Sigma^{+}$
electronic system into 12 separate files.  Since absorption
bands with the same $\Delta{v}$ strongly overlap in wavenumber,
the computed spectrum has the appearance of 6 bands for $\Delta v$ = -3,
-2, -1, 0, 1 and 2.  Each line list file 
contains a listing for each line (sorted by branch) with the
values of $v^{\prime}$ and $v^{\prime\prime}$ for the two states, the
rotational quantum number $J^{\prime\prime}$ of
the lower state, and an index indicating if the transition belongs to a P,
Q, or R-branch.  (These data are now available in electronic form at
http://bernath.uwaterloo.ca/CrH .)
Thus, the upper rotational quantum number $J^{\prime}$, the
wavenumber of the transition, the Einstein $A$ coefficient of the
transition, and the excitation energy (term value) in cm$^{-1}$ of the lower state can be
found. Both electronic states are sextets so each rotational level (N quantum
number) is split (by coupling with the net electron spin of 5/2) 
into six energy levels: F$_1$, F$_2$, F$_3$, F$_4$, F$_5$ and F$_6$
with J=N+5/2, N+3/2, N+1/2, N-1/2, N-3/2, N-5/2,  respectively. 
Indexes (1 to 6) for these spin components
are therefore provided for the upper and lower states and are used to label the
branches.  The $e/f$ parity of the lower state is also provided, although
we do not need it for the opacity calculations.  A total of $\sim$14,000
lines (not including isotopes; \S\ref{opac}) are found in the resulting
line list.  This is $\sim$5 times the size of previous CrH compilations,
but is still considerably smaller than the $\sim$300,000,000 lines in the  
current line list for the triatomic molecule H$_2$O (Partridge and Schwenke 1997).

\section{The Calculation of CrH Opacities}
\label{opac}

In order to calculate spectral models of brown dwarfs and giant planets,
opacities are required. However, the computer time and memory necessary to calculate
the contribution to the overall opacity of a large number of molecular lines
often precludes (but does not render impossible) calculating detailed opacities 
during a model atmosphere calculation at each temperature and pressure point.
Given this, we prefer to precompute the absorption cross section of CrH and other
molecules at a large number of temperature ($T$) and gas pressure ($P$)
points regularly arranged in a table, so that during a model calculation the cross section of
species such as CrH is calculated by interpolation in that table.  For each
spectroscopically active molecule, a tabulated $T/P$ point has an 
associated file that contains the monochromatic absorption
cross section in cm$^2\,$ molecule$^{-1}$ calculated at one 
wavenumber intervals over a wide, but fixed spectral range.
No averaging is done.
The cross section of that molecule at an arbitrary $T/P$ point is obtained
by interpolating between the four closest tabulated values of temperature and
pressure. Given the abundance of the molecule, its contribution to the total
opacity at a general $T/P$ point is then calculated.

Since the Einstein $A$ coefficient includes all the details of the intrinsic
band strength, together with the Franck-Condon and H\"onl-London factors, we
do not need to consider these quantities separately.  Without considering
separate isotopic versions of CrH, only the Einstein coefficient, the lower rotational quantum
number $J^{\prime\prime}$, the transition wavenumber ($\bar{\nu}$), the
term value of the lower state $F^{\prime\prime}$, and the 
internal partition function $Q$ are required to
obtain the frequency-integrated cross section (line strength):
\begin{equation}
S = \frac{1}{8\pi\bar{\nu}^2} A(2J^{\prime\prime}+1)
\exp(-F^{\prime\prime}hc/kT) (1-\exp(-hc\bar{\nu}/kT))/Q\, .
\label{eqS}
\end{equation}
$F^{\prime\prime}$ is in wavenumbers and the stimulated emission
factor is included.

Given the vibrational and rotational constants $w_e$, $w_ex_e$, $B_e$ and
$D_e$ for the X and A electronic states (Bauschlicher et al. 2001), the
vibrational and rotational contributions to the individual partition
functions of the X and A states are calculated using asymptotic formulae from Kassel
(1933a,b). The total internal partition function is calculated from

\begin{equation}
Q = 6(Q_r^{\prime\prime}Q_v^{\prime\prime} +
Q_r^{\prime}Q_v^{\prime}\exp(-T^{\prime}hc/kT))\, ,
\label{eqQ}
\end{equation}
where $Q_r^{\prime\prime}$ and $Q_v^{\prime\prime}$ are, respectively, the
rotational and vibrational partition functions of the X state, $Q_r^{\prime}$ and $Q_v^{\prime}$ are
the corresponding partition functions of the excited A state, $T^{\prime}$ is the excitation
energy of the A state (in wavenumbers), and the factor of six accounts for the
multiplicity (the electronic contribution to the partition function).
There are, in fact, several other states predicted to lie near the
A$^6\Sigma^{+}$ state (Ram, Jarman, and Bernath 1993; Dai and
Balasubramanian 1993), but their contribution to the sum in eq. (\ref{eqQ}) at the temperatures of
interest in brown dwarfs is negligible.

We include, along with the most abundant isotope of chromium ($^{52}$Cr, which makes up 83.8\% of
chromium), the contributions of $^{50}$Cr, $^{53}$Cr, and $^{54}$Cr,
which make up 4.4\%, 9.5\% and 2.4\%, respectively. Given the reduced mass of the most
abundant isotopic form $^{52}$Cr$^{1}$H, and the reduced mass of any
isotopically substituted form of CrH, the vibrational, rotational, and
vibration-rotation coupling constants can be found from the
corresponding constants for $^{52}$Cr$^{1}$H by multipling by the square root of the reduced
mass ratio raised to an appropriate integer power (Herzberg 1950).  By
identifying the lower and upper levels involved in a given transition,
together with the vibrational and rotational quantum numbers, the shifts in the
levels can be calculated approximately. Hence, the displacement in the transition
wavenumber is obtained for a given isotopic version of CrH.

However, we do not correct for the slightly different partition functions of
the isotopic versions, nor do we allow for the small
change in the line strengths due to the shift of the lower energy level
(hence, its Boltzmann factor).  We also ignore the effect of changes in the line frequency
on its oscillator strength, and the change in the Franck-Condon factors due to the
small changes in the wavefunctions.  We think, however, that these effects
are very small compared with the shifting of the line positions and the
consequent ``filling in'' of the absorption between the lines of the main isotopomer.
Given eq. (\ref{eqS}) above, the line
strength for each isotopic version is obtained by multiplying it by the
fractional abundance of that isotope.  All 12 files are read and the lines
are calculated by cycling through the data for each of the four isotopes.

Finally, in order to calculate the absorption cross section per molecule on 
a frequency grid the integrated line strengths are multiplied by
a suitably broadened profile.  Our broadening algorithm is an extension of
that employed for FeH by R. Freedman (private communication) based on a Lorentzian
lineshape function. For each line given in an input file, the
parameter $W_L$ is given (as provided by R. Freedman), which is the line
full width in cm$^{-1}$ at 296 K and 1 atmosphere.
At other temperatures and pressures a scale factor of $(296 \,{\rm
K}/T)^{w_x}(P_{H_2}/1 \,atm.)$ is used,
where $w_x$ is 0.7.  $W_L$ is generally $\sim$0.045 cm$^{-1}$.

This whole process is repeated for all four isotopic forms of CrH at all the
tabular $T/P$ points.
The opacities at a given general $T/P$ point are obtained by interpolation in these files.

\section{Representative CrH Opacity Plots}
\label{opacplots}

Using the procedures described in \S\ref{opac}, we have created a program
for generating CrH cross sections for the six resulting band sequences with band heads from
$\sim$0.7 \mic to $\sim$ 1.4 \mic.  Readers can obtain tables generated by this
program from the first author.  Figure \ref{fig:1} depicts the resulting CrH cross
sections for representative pressures
and temperatures at which CrH is typically found in substellar atmospheres.
Specifically, opacity spectra for $T/P$ pairs of 1500 K/10 bars (red), 2000
K/10 bars (green), and 2000 K/100 bars (blue) are portrayed.   The
corresponding opacities using the older database for the
0--0 transition of CrH of Ram, Jarman, and Bernath (1993), as calculated by R. Freedman (private communication), 
are two orders of magnitude weaker.   The lion's share of the difference
between the old and the new CrH opacities can be traced to an increase by a factor of $\sim$13.5
in the oscillator strength and to a previously inappropriate division by 6, 
the electronic spin degeneracy factor.  The A--X 0--0 
band at $\sim$0.86 microns (just shortward of the
neighboring FeH feature at $\sim$0.87 microns) and the A--X 0--1 CrH band 
near $\sim$0.997 microns (just longward of the classic Wing-Ford band of FeH)
are prominent and, in principle, diagnostic features in measured L dwarf
spectra (Kirkpatrick et al. 1999a,b, 2000).
Importantly, Kirkpatrick et al. (1999b, 2000) demonstrate that the strengths
of these two CrH bands peak near a subtype of L5-L6 and that at later subtypes
(which are in principle at lower \teffs) they gain ascendancy over the
associated FeH features. As is clear from a comparison of the red and green or blue curves,
away from the band heads the CrH opacity is an increasing function of temperature.  Furthermore, and as
expected, higher pressures result in greater line overlap and, hence, a smoother opacity spectrum.

Some other molecules that are important opacity sources in the atmospheres of substellar mass objects
include H$_2$O, H$_2$, CO, CH$_4$, TiO, and VO.   To provide the context of the new CrH
opacities, we portray in Fig. \ref{fig:2} the CrH, H$_2$O and VO per-molecule cross sections at 1500 K
and 10 bars.  In atmosphere calculations, the abundances are also germane, but Fig. \ref{fig:2} indicates
the different wavelength regions for which each of these species is important.  As Fig. \ref{fig:2} indicates,
CrH can be prominent shortward of one micron, particularly after the higher-temperature species TiO and VO 
recede from view.

\section{The Inferred CrH Abundance in an L Dwarf Atmosphere}
\label{bdspec}

Theoretically, both CrH and FeH should be  
important as \teff decreases through 2000 K, reaches the main-sequence edge
(\teff$\sim$1600-1750 K; Burrows et al. 2001),    
and dives into the brown dwarf realm.
Hence, with \teffs for L dwarfs of from
$\sim$2300 K to $\sim$1300 K (a range which straddles the stellar-substellar boundary),
CrH should be particularly relevant in their atmospheres (Pavlenko 2001; Kirkpatrick \etal 1999a,b).

Chemical equilibrium studies indicate that CrH is expected to survive to lower
temperatures in the atmosphere of an L or brown dwarf than FeH,  
perhaps reaching temperature levels of $\sim$1400--1600 K,
as opposed to $\sim$1500--2000 K for FeH.  However, the actual CrH abundance in an atmosphere
will depend not only on CrH chemistry, but in principle on the effects of rainout, 
settling, and depletion in the dwarf's gravitational field (Burrows and Sharp 1999; Lodders 1999)
and on non-equilibrium dynamics in the dwarf's convective zone.
These phenomena can be difficult to model.
Fortunately, with better CrH opacities 
one can use L dwarf spectra to constrain the actual CrH  
abundances in L dwarf atmospheres.  
A similar study was recently carried out for the CO molecule 
by Noll, Marley, and Geballe 1997, who 
determined that the CO abundance in the upper atmosphere of 
Gliese 229B is far from its chemical equilibrium value.   

In the past, in a reversal of the above philosophy, 
astronomers have occasionally used stellar spectra to infer the opacity in a spectral band of
an exotic, but important, molecule under non-terrestrial conditions.  
This was the tradition for TiO and VO in the early days of M dwarf studies
(Mould 1976) and has been attempted recently for FeH (Schiavon, Barbuy, and Singh 1997).
However, this approach does not allow an independent extraction of the distribution
and abundance of the molecule and can lead to very significant errors 
in the inferred cross sections.

To derive the CrH abundance in a representative L dwarf 
using the new opacities, we have calculated a few
synthetic L dwarf spectra, using the approach of Burrows \etal (2002).  Though the presence of clouds
complicates the modeling effort (Burrows \etal 2001), we 
reach a few preliminary conclusions about both CrH abundances 
in L dwarfs and the population of L dwarf grains.  
We modeled clouds which consisted of particles of forsterite (Mg$_2$SiO$_4$)
with an assumed modal particle size of 50 \mic (Cooper \etal 2002).  The cloud top 
(for a semi-infinite cloud) was put at an atmospheric
temperature level of 1700 K.  Above this temperature, all the magnesium was
assumed to reside in forsterite.  The cloud optical properties 
were calculated using Mie theory for spherical particles. 

In Fig. \ref{fig:3}, we plot three spectral models from 0.75 \mic to 1.05 \mic
at \teff = 1700 K and for a gravity of 10$^{5.5}$ cm s$^{-2}$, 
along with the measured spectrum of the L5 dwarf 2MASSI J1507038-151648 (black curve, 2MASS-1507)
(Kirkpatrick et al. 1999).  This L dwarf should be near the main sequence edge.
Clearly seen in the 2MASS-1507 data are the two major CrH bands
near 0.86 \mic and 1.0 \mic.   (Note that the theoretical FeH opacities
are currently undergoing a major revision and that we are using for these calculations 
the old set, now known to be substantially off (P. Bernath \etal, in preparation).
Hence, we make no attempt to fit the FeH features at $\sim0.87$ \mic and $\sim0.99$ \mic,
each of which is adjacent to a CrH feature.)

The green curve in Fig. \ref{fig:3} is a model (model C on the figure) 
without grains and has a CrH/H$_2$ number ratio of $2\times 10^{-9}$ above 1400 K.
This curve is clearly too steep to explain the spectrum of 2MASS-1507 shortward of one micron.
The red and blue curves (models A and B on Fig. \ref{fig:3}) include 
the forsterite cloud with 50 \mic particles and have CrH/H$_2$ abundance ratios of 
$4\times 10^{-9}$ and $2\times 10^{-9}$, respectively.  These values seem to
bracket the 2MASS-1507 data.  Shortward of 0.85 \mic, the cloudy models fit much better
than the cloud-free model, but are too flat longward of 0.95 \mic.  
Presumably, this flatness at longer wavelengths is a consequence 
of our imperfect cloud model and the assumed wavelength dependence of
the real and imaginary indices of refraction of the grains (Scott and Duley 1996).   
For a given \teff, since clouds suppress flux in the $Z$ and $J$ bands (not shown), radiation
must come out somewhere else.  It does so in just the right region shortward of $\sim$0.85 \mic to 
compensate for the flux deficit of the cloud-free model there (Fig. \ref{fig:3}).  Hence, for mid-L
spectral subtypes, the combination of K I wing opacity (Burrows, 
Marley, and Sharp 2000) with our forsterite cloud opacity
can account approximately for what is seen shortward of 0.95 \mic.  
Moreover, the cloud decreases the depth of the 0.93 
\mic water feature (red [A] or blue [B] model versus green [C] model),
in accord with the observations (black).  For 
T dwarfs in the optical (Burrows \etal 2002), the strong effect
of clouds is missing and the corresponding slope is steeper and dominated 
by the red wing of the 7700 \AA\ feature of K I alone.  This slope change is characteristic
of the L$\rightarrow$T transition.   For 2MASS-1507, we are not able
to obtain even crude fits in the optical or 2MASS-1507's $J-K$ color near 
1.41 magnitudes (Dahn \etal 2002) without clouds or for clouds with small particles 
(e.g., $\sim$0.1 \mic <$<$1.0 \mic).  Nevertheless, our cloud model is primitive.    
Fortunately, the inferred CrH abundance is not a strong function of cloud 
properties and abundances three times higher or lower (not shown) are clearly 
excluded.  We conclude that the CrH/H$_2$ number ratio in 2MASS-1507 is probably in the 
$\sim 2-4\times 10^{-9}$ range.  The Anders and Grevesse (1989) elemental abundance of 
chromium is $\sim 4.4\times 10^{-7}$ by number, which results in an atmospheric CrH/Cr number ratio
of $\sim 2.3-4.5 \times 10^{-3}$ above $\sim$1400 K, 
in reasonable agreement with chemical abundance calculations
(M. Marley, R. Freedman, and K. Lodders, private communication).

\section{Conclusions}
\label{conclusions}

Using the vibrational
and rotational constants for the A$^6\Sigma^{+}$ state of the CrH molecule 
recently derived by Bauschlicher et al. (2001), we have calculated new line
lists and opacities for the 12 bands of its A$^6\Sigma^{+}$ -- X$^6\Sigma^{+}$ 
transitions.  The resulting theoretical data are available from the authors.
CrH is a defining molecule of the L dwarf spectroscopic class and accurate
opacities as a function of temperature and pressure are necessary for 
spectral syntheses and to extract CrH abundances for L dwarf atmospheres.
In a tentative first step, we use the new theoretical data to obtain such 
an abundance for the L5 dwarf 2MASSI J1507038-151648. The CrH/H$_2$ number ratio
we find is $\sim 2-4\times 10^{-9}$, in reasonable agreement with expectations.

\acknowledgments

This work was supported in part by NASA under grants
NAG5-10760, NAG5-10629, NAG5-7499, and NAG5-7073. Support
was also provided by the NASA Laboratory Physics Program and the
Natural Sciences and Engineering Research Council of Canada.
The authors would like to thank Richard Freedman for exploring with
a parallel set of calculations the consequences of the new line
lists for the resultant CrH opacities, M. Dulick for providing the
expressions for the H\"onl-London factors, and David Sudarsky for helping
with the calculations of cloud properties.  In addition, they would
like to thank Ivan Hubeny for consultations on radiative transfer technique.
The new CrH line lists are available at http://bernath.uwaterloo.ca/CrH 
and CrH opacity tables in a fixed format can be
obtained from the first author at burrows@as.arizona.edu.

\clearpage

% figure 1
\figcaption{The logarithm (base ten) of the absorption cross section of CrH versus
wavelength (in microns) from $\sim$0.7 $\mu$m
to 1.5 $\mu$m, at various temperatures and pressures.  The blue curve at 100
bars and 2000 K depicts the effect of large pressure broadening (when compared with the red curve 
at 10 bars and 1500 K).  A comparison of the green curve (10 bars, 2000 K)
with the red curve (10 bars, 1500 K) portrays the effect of increasing temperature.  The A--X 0-0 band is 
the strongest and is the third from the left near $\sim$0.9 $\mu$m.
As the temperature and pressure decrease the cross section range in a given
band widens and executes larger variation with wavelength. 
\label{fig:1}}

% figure 2
\figcaption{The logarithm (base ten) of the absorption cross sections of CrH, H$_2$O, and VO versus
wavelength (in microns) from $\sim$0.3 $\mu$m
to 3.5 $\mu$m, at 1500 K and 10 bars.  The blue curve with the strong ``blue" slope 
is for VO, a species which is important above $\sim$1800 K.  The red curve is for H$_2$O,
whose opacity rises towards the near infrared.  The green curve is for CrH (as in Fig. \ref{fig:1}),
which comes into its own in very cool atmospheres between 2000 K and 1400 K.
\label{fig:2}}

% figure 3
\figcaption{The log (base ten) of the absolute flux density (${\cal F}_\nu$) in milliJanskys versus wavelength
($\lambda$) in microns from 0.75 \mic to 1.05 \mic for self-consistent
theoretical solar-metallicity models of the L5 dwarf 2MASS-1507.  
Also included are the corresponding data (in black) for 2MASS-1507 
from Kirkpatrick \etal (1999b).  All models are for \teff = 1700 K and a gravity of $10^{5.5}$ cm s$^{-2}$. 
The dashed red line depicts a model (A) with a forsterite cloud and a CrH/H$_2$ number abundance ratio of $4\times 10^{-9}$,
the blue line depicts a model (B) with a forsterite cloud and a CrH/H$_2$ number abundance ratio of $2\times 10^{-9}$, 
and the green line depicts a cloud-free model (C) with a CrH/H$_2$ number abundance ratio of $2\times 10^{-9}$.
Below temperatures of 1400 K, the CrH abundance was set to zero.
Indicated with arrows are the positions of the CrH, FeH, H$_2$O, and K I (7700\AA) features in this
spectral range.  Also prominent are the Cs I lines at 8523 \AA\ and 8946 \AA, the Na I line at 8195 \AA,   
and the Rb I lines at 7802 \AA\ and 7949 \AA.  These spectra
have been deresolved to an $R$($\lambda/{\Delta\lambda}$) of 1000.
\label{fig:3}}

\end{document}